\documentstyle[12pt]{article}
\input epsf

\begin{document}

\begin{flushright}
\vbox{\begin{tabular}{l} MADPH-98-1056  \\
	FERMILAB-PUB-98/128-T \\
	April 1998
	\end{tabular}}
\end{flushright}

\begin{center}
\vspace{+1cm}
\Large
{\bf Excited glue and the vibrating flux tube} \\
\vskip 0.5cm
\large
Theodore J. Allen \ {\it and\/} \ M. G. Olsson \\
{\small \em Department of Physics, University of Wisconsin, \\
1150 University Avenue, Madison,
	\rm WI 53706} \\
\vspace*{+0.4cm}
	Sini\v{s}a Veseli \\
\vskip 0.1cm
{\small \em Fermi National Accelerator Laboratory, \\ P.O. Box 500, Batavia,
	\rm IL 60510}
\end{center}
\thispagestyle{empty}
\vskip 0.7cm

\begin{abstract}
Recent lattice results for the energy of gluonic excitations as a function
of quark separation are shown to correspond to transverse relativistic
flux tube vibration modes. For large quark separations all states appear to
degenerate into a few categories which are predicted uniquely, given the
ground state.
\end{abstract}

\newpage

\section*{Introduction}

Mesons in which the gluons are in an excited state have been discussed for
some time.  There are two main pictures that have evolved for treating
these excited states.  The first is the constituent gluon approach where
the quarks and a gluon move in an MIT bag \cite{bag} or a potential model
\cite{potential,swansonszczepaniak}.  The second picture envisions the
quarks to be connected by a string or flux tube
\cite{isgurpaton,string,perantonismichael} which has quantized transverse
vibrations.  In this case the flux tube can be thought of as a coherent
gluonic state.  In all of these models the resulting meson is analogous to
the diatomic molecule where the gluonic degrees of freedom are the
``electronic state'' that can assume many levels of excitation.  Each
excited state yields an interaction energy that acts as an adiabatic
potential in which the quarks or ``ions'' move.  The ground state of the
glue corresponds to standard meson states and the excited glue to ``hybrid
meson'' states.

Recently the excited glue states with fixed end points have been
investigated in detail by lattice simulation \cite{lattice}.  These
calculations are done with an improved action in the quenched approximation
for a variety of gluonic operators, and on several anisotropic lattices.
It is our purpose here to point out that the systematics of the gluon
states are extremely simple from the vibrating relativistic flux tube point
of view.  To a remarkable extent the gluon states group themselves into a
few highly degenerate states at large quark separations, reflecting the
well known degeneracy of the quantized two-dimensional harmonic oscillator.

We further show that, given the ground state potential, the hybrid
adiabatic potentials are uniquely predicted and agree well with the lattice
results.  Our calculation is fully relativistic and does not introduce
arbitrary procedures as required by previous work \cite{isgurpaton}.

\section*{Relativistic strings}

It is a common misconception that a free relativistic string {\it must} be
formulated in twenty six dimensions in order to be consistently quantized.
In fact, quantized theories of a single non-interacting relativistic string
can be defined consistently in any spacetime dimension smaller than twenty
six using the standard string theoretic methods. Long ago, Brower and
Goddard and Thorn \cite{browergoddardthorn} showed that free bosonic string
theories in spacetime dimensions $D\leq 26$ are free of ghost (negative
norm) states as long as the first excited state is not tachyonic.
Subsequently Rohrlich \cite{rohrlich} found an oscillator quantization of
the non-interacting relativistic string that is manifestly free of ghosts
in any dimension, while Polyakov \cite{polyakov} quantized the string as a
sum over random surfaces that is consistent in dimensions twenty-six or
smaller. It is only in the context of dual models and their superstring
offspring that the theory becomes consistent in a single (critical)
dimension.  This is because unphysical states that may be consistently
eliminated from a free string cannot be consistently eliminated from an
interacting string theory \cite{GSW}.

The Nambu-Goto action, with fixed end boundary conditions may be quantized
consistently in $D=4$ using the Gupta-Bleuler method in the temporal gauge,
$X^0(\sigma,\tau) = t = {P^0\over \pi a}\tau$.  The energy of a string of
tension $a$ and distance $r$ between the fixed ends,
\begin{eqnarray}
\label{energies}
E(r) &=& \sqrt{ (ar)^2 + 2 \pi a (N + c )}\ , \\
\label{level}
N &=& \sum_{m=1}^\infty m(n^+_m + n^-_m),
\end{eqnarray}
follows from the zero mode of the Virasoro constraint
\cite{nesterenko,AOV}.  The index $m$ labels the mode level which is
occupied by $n^+_m$ phonons of positive helicity and $n^-_m$ phonons of
negative helicity.  The constant $c$ is an arbitrary normal ordering
constant subject only to the constraint, $c \geq -1$, of the no-ghost
theorem.

In the temporal gauge, Lorentz invariance does not impose any restriction
on spacetime dimension $D$ \cite{rohrlich} as it does in light-cone gauge
and only requires that $c$ be chosen such that the system has a rest frame.
This formally infinite constant is often calculated
\cite{brinknielsen,lambiasenesterenko,alvarezarvis} by summing the Casimir
zero-point energies using zeta function regularization, yielding the value
$c= -{D-2\over 24}$.  The standard BRST quantization method also yields the
values $D=26$ and $c=-1$. The Gupta-Bleuler method we use here yields a
consistent quantum theory of a single string for any value of $c\geq 0$.
It is well known that different methods of quantizing theories with
constraints, such as the Nambu-Goto string, need not be equivalent and may
differ in their energy spectra as well as in their dynamical degrees of
freedom.

The excited glue states are completely specified by the separation between
the ends and the occupation number of each of the modes, $|\psi\rangle =
|\, {\bf r},\{n_m^+,n_m^-\}\rangle$.  These states can be labeled in
molecular notation \cite{landaulifshitz} by three quantities.  The first is
their angular momentum along the quark axis,
\begin{equation}
\Lambda = \sum_{m=1}^\infty (n^+_m - n^-_m).
\end{equation}
States with $|\Lambda| = 0,1,2,\ldots$ are denoted $\Sigma, \Pi, \Delta,
\ldots$.  The second label is the CP value of the glue which appears as
either a subscript $g$ or $u$ depending on whether CP is even or odd
respectively.  The $\Sigma$ states ($\Lambda = 0$) are labeled additionally
by a superscript $\pm$ denoting their parity under reflection through a
plane in which the axis lies. The CP value of the flux tube is determined
by \cite{isgurpaton}
\begin{equation}
\eta_{CP} = (-1)^N .
\end{equation}
In the vibrating flux tube model the $\Sigma^+$ and $\Sigma^-$ states are
always degenerate as well.

For $N=0$ and $1$ the glue states are uniquely $\Sigma_g^+$ and $\Pi_u$
respectively.  For $N=2$ the flux tube can be excited in the $\Sigma_g$,
$\Pi_g$, and $\Delta_g$ states.  For an arbitrary $N>1$, states with
$\Lambda=0,1,\ldots,N$ can be excited.  This degeneracy is a firm
prediction of the flux tube picture and is expected to hold for large
$Q\bar{Q}$ separations.

\section*{Testing the flux tube vibration picture}

In this section we compare the predictions of Eq.~(\ref{energies}) for flux
tube vibrations with the results of a lattice simulation of quenched QCD
\cite{lattice}.  The lattice energies are given relative to the ground
state ($\Sigma_g^+$) energy at a quark separation of $r=2r_0$, where $r_0$
is a hadronic scale distance determined \cite{MP} from the $\Sigma_g^+$
data at large $r$.

The ground state ($\Sigma_g^+$) potential from Eq.~(\ref{energies}) with
$N=0$, together with a short distance $Q\bar{Q}$ color singlet potential
$-4\alpha_s/3r$ and an additive constant, gives
\begin{equation}
\label{groundpotential}
V_{\Sigma_g^+}(r) = -{4\alpha_s\over 3r}+\sqrt{a^2 r^2 + 2\pi a c}+C \ .
\end{equation}
The existence of a potential for all values of $r$ requires $c$ non-negative.
If $c=0$, the ``Cornell'' potential is recovered,
\begin{equation}
\label{cornellpotential}
V_{\Sigma_g^+}(r)\, {\buildrel {c\rightarrow 0} \over \longrightarrow }\,
 V_{\rm Cornell} =\ -{4\alpha_s\over 3r} + C + a \, r \ .
\end{equation}
The best fit of Eq.~(\ref{groundpotential}) to the lattice data occurs in
fact where $c=0$.  Assuming this, we obtain for the best representation of
$\Sigma_g^+$
\begin{eqnarray}
\label{rnaught}
\alpha_s &=& 0.234\ , \nonumber \\
a r_0^2 &=& 1.34 \ , \nonumber \\
C r_0 &=& -2.51 \ , \\
c &=& 0. \nonumber
\end{eqnarray}
Once the above values have been fixed by the ground state lattice data, all
of the excited states are predicted uniquely, with
\begin{equation}\label{higherN}
V_N(r) = \sqrt{(ar)^2 + 2\pi a (N + c)} + C 
\end{equation}

The gluonic excitation energies obtained by lattice simulation are shown in
Figs.~1--3 in molecular notation.  In Fig.~1 the $\Lambda=0$ states
$\Sigma_{g,u}^{\pm}$ are plotted.  The lowest $\Lambda = 1$ ($\Pi_{g,u}$)
and $\Lambda=2$ ($\Delta_{g,u}$) states are displayed in Figs.~2 and 3
respectively.  For $\Lambda \neq 0$, the $+\Lambda$ and $-\Lambda$ states
are degenerate.

In the figures we also show the $N = 1$, $2$, and $3$ predictions superimposed
upon the $\Sigma$, $\Pi$, and $\Delta$ lattice data \cite{lattice}.  We
observe general agreement at large $r$.  The $\Pi$ and $\Delta$ states
follow the vibrating string prediction even down to small $Q\bar{Q}$
separations.  The $\Sigma^-$ states do not closely resemble either the $N=2$
or $3$ curves.  One possible explanation is that they have admixtures of
higher $N$ excitations.

Because the flux tube may be viewed as a coherent gluonic state, one might
expect that for small quark separation it should continue smoothly to the
perturbative configuration of one gluon and a $Q\bar{Q}$ color octet.  We
test this assumption by modifying the prediction of
Eq.~(\ref{higherN}) by adding to the excited potentials a color octet short
distance repulsion,
\begin{equation}
\label{octet}
V_{\rm octet}(r) = {\alpha_s\over 6 r}\ ,
\end{equation}
where $\alpha_s$ was determined (\ref{rnaught}) by the short distance color
singlet attraction of the $\Sigma_g^+$ ground state.  Our results (dashed
curves) indicate that this short range behavior seems to improve most of
the predictions, if only marginally.

\section*{Conclusions}

Relativistic vibrating string models have been widely discussed in the
literature. Some authors maintain that a tachyonic ground state ($c < 0$)
is required in fewer than 26 dimensions
\cite{lambiasenesterenko,alvarezarvis,polchinskistrominger,dienescudell}.
Others maintain that there are quantization ambiguities that allow, at
least for a non-interacting string, consistent quantization in four
dimensions \cite{browergoddardthorn, rohrlich, nesterenko, patrascioiu}.
We advocate the later interpretation and show that the vibrating (phonon)
excitations closely correspond to a lattice simulation of QCD.

The success of this simple flux tube picture strongly supports the string
vibration modes as being the correct low energy degrees of freedom for the
gluonic excitations in hybrid mesons.  The original work of Isgur and
Paton \cite{isgurpaton} has the same large $r$ limit as our result but is
intrinsically non-relativistic.  Since waves propagate with the speed of
light on a flux tube, a relativistic model is mandated.  Furthermore, the
Isgur-Paton model introduces an arbitrary radial functional dependence to
avoid unphysical behavior at $r=0$.  In this function a free parameter can
be tuned so that the potential has zero slope at $r=0$ for the $N=1$ state.
For larger $N$ the intercept rises as required, but now the slope becomes
progressively more negative.  This effect is demonstrated numerically in
\cite{swansonszczepaniak}.  Our model predicts a unique potential for the
excited states once the parameters are determined by the ground state.

\section*{Acknowledgements}

We thank C.J. Morningstar for providing us the data of reference
\cite{lattice}.  This work was supported in part by the US Department of
Energy under Contracts Nos.~DE-AC02-76CH03000 and DE-FG02-95ER40896.

\epsfxsize=6 truein

\begin{figure}
\hbox{\hskip 0 in \epsfbox{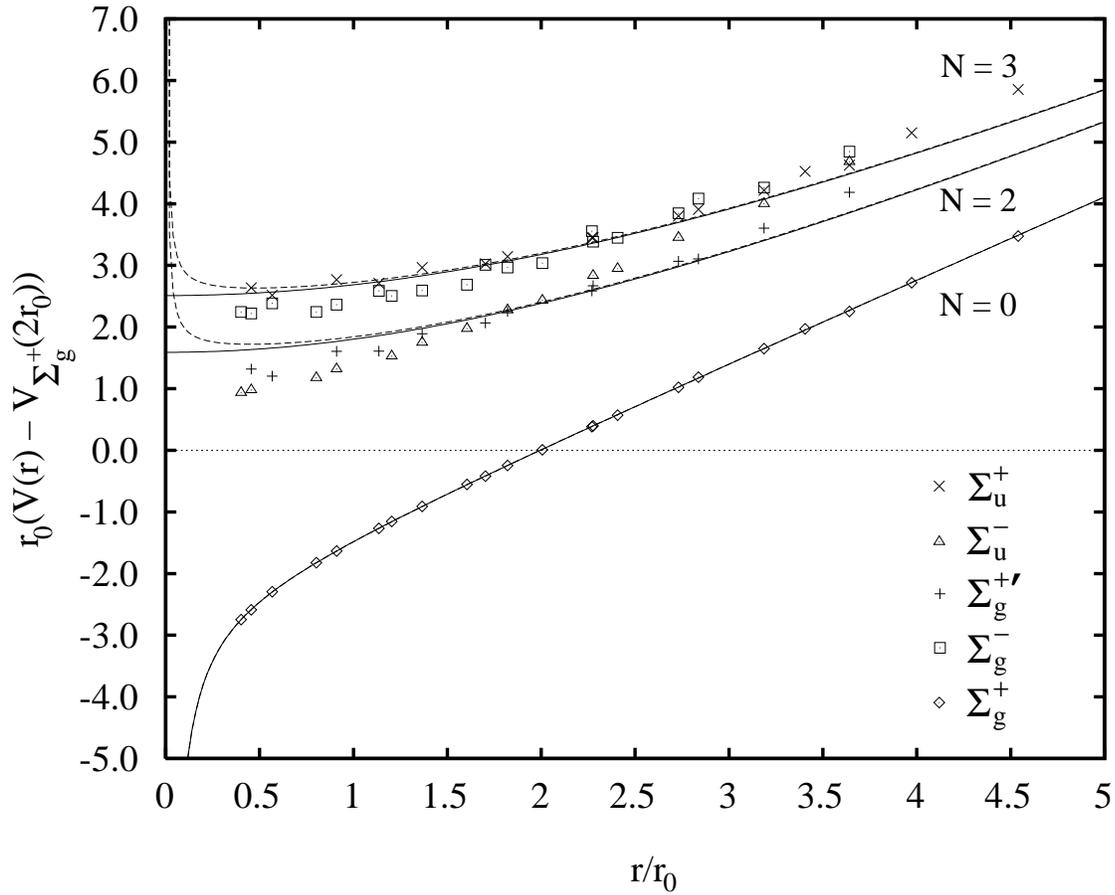}}
\caption{The sigma states.  The bottom line is the best fit of
Eq.~(\protect{\ref{groundpotential}}) to the $\Sigma_g^+$ states, which
completely determines the other curves.  The other solid lines are the
predictions from the vibrating flux tube, while the dashed lines are the
predictions of the flux tube plus an octet repulsive potential.  The
$\Sigma_g$ states correspond to even $N$, while the $\Sigma_u$ correspond
to odd $N$.}
\end{figure}

\epsfxsize=6 truein

\begin{figure}
\hbox{\hskip 0 in \epsfbox{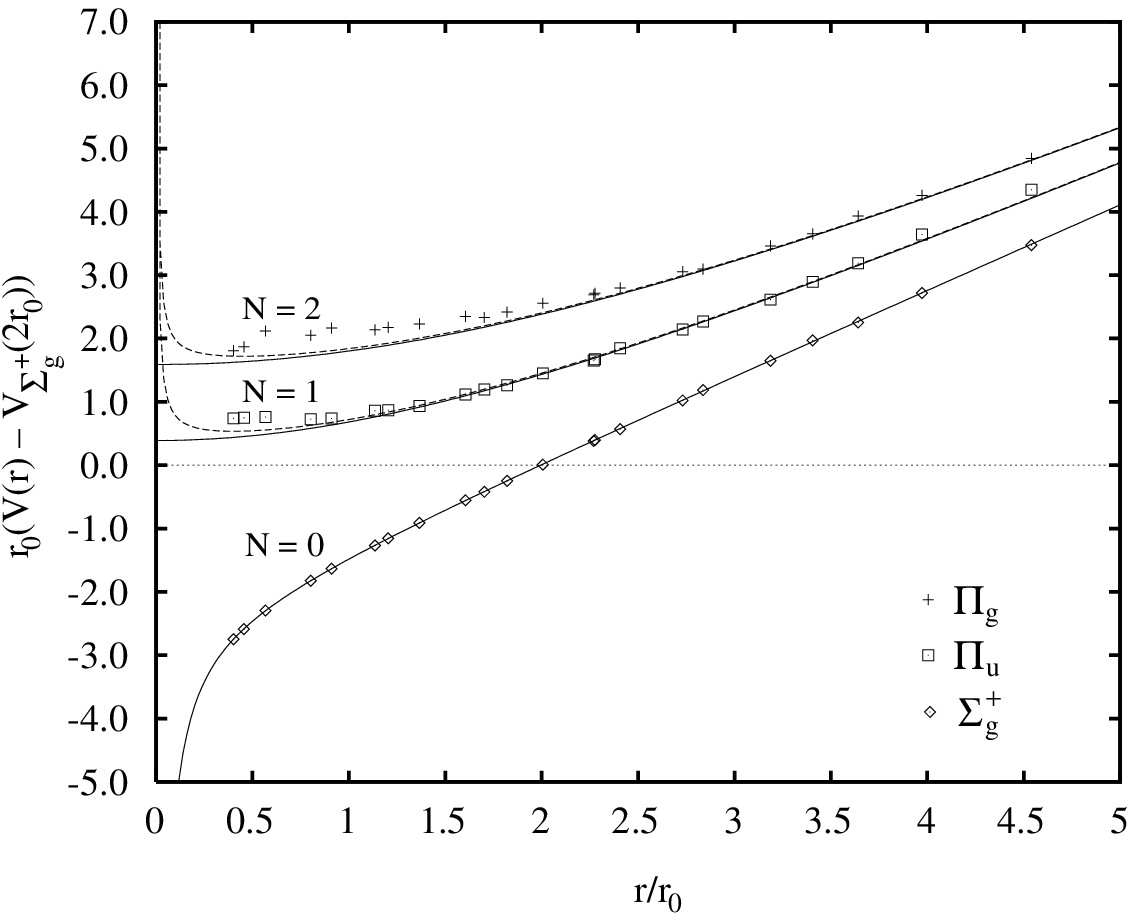}}
\caption{The pi states.  The bottom line is the best fit of
Eq.~(\protect{\ref{groundpotential}}) to the $\Sigma_g^+$ states, which
completely determines the other curves.  The other solid lines are the
predictions from the vibrating flux tube, while the dashed lines are the
predictions of the flux tube plus an octet repulsive potential.}
\end{figure}

\epsfxsize=6 truein

\begin{figure}
\hbox{\hskip 0 in \epsfbox{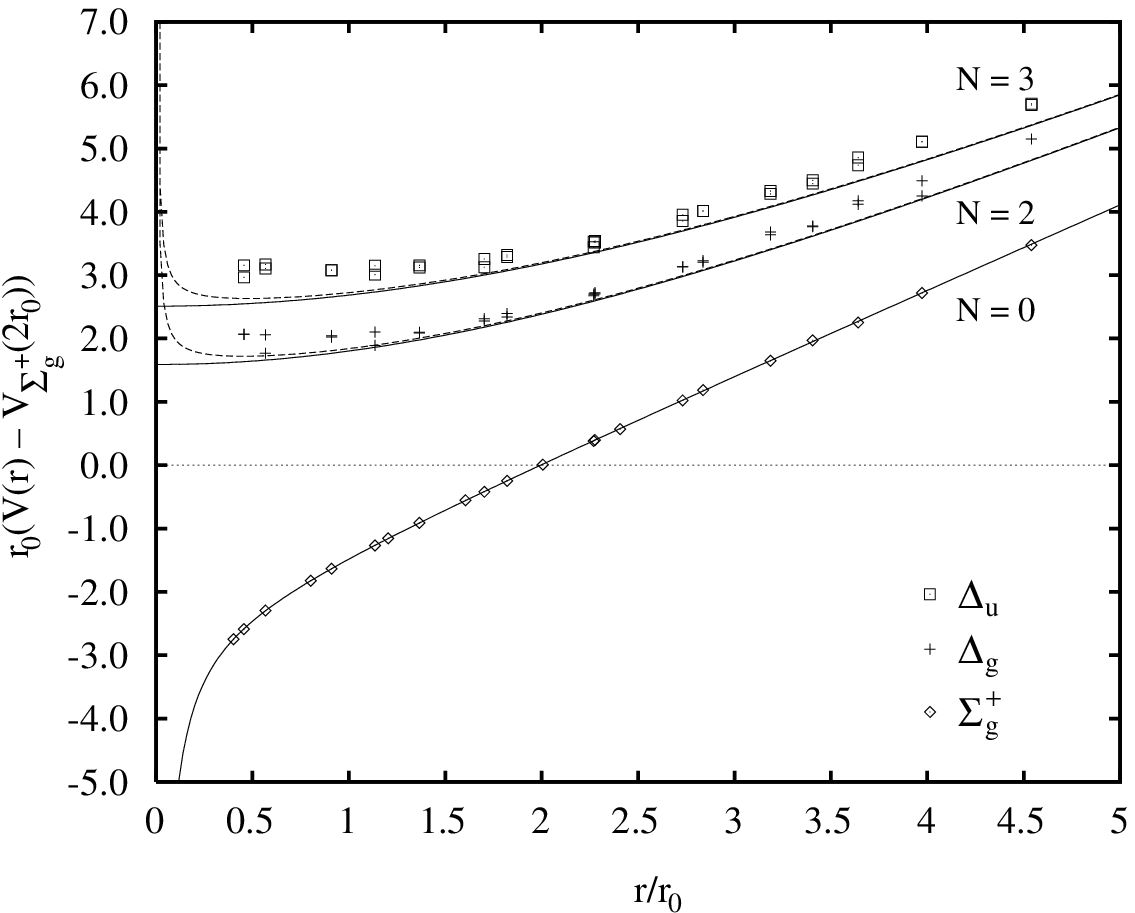}}
\caption{The delta states.  The bottom line is the best fit of
Eq.~(\protect{\ref{groundpotential}}) to the $\Sigma_g^+$ states, which
completely determines the other curves.  The other solid lines are the
predictions from the vibrating flux tube, while the dashed lines are the
predictions of the flux tube plus an octet repulsive potential.}
\end{figure}

\end{document}